\documentclass[reprint,aps,prd,superscriptaddress,showkeys,showpacs]{revtex4-1}
\usepackage{epsfig,amsmath,natbib}

\usepackage{aas_macros}
\usepackage{amssymb}
\usepackage{amsmath}
\usepackage{dsfont}
\usepackage{hyperref}
\usepackage{color}
\usepackage{pbox}
\usepackage{booktabs}
\usepackage[dvipsnames]{xcolor}

\hypersetup{
	colorlinks=false,
	citecolor=green
}

\newcommand{\msun}{{\rm M}_\odot}

\newcommand{\q}{q_s}

\newcommand\lsim{\mathrel{\rlap{\lower4pt\hbox{\hskip1pt$\sim$}}
        \raise1pt\hbox{$<$}}}
\newcommand\gsim{\mathrel{\rlap{\lower4pt\hbox{\hskip1pt$\sim$}}
        \raise1pt\hbox{$>$}}}

\graphicspath{{figures/}}

\begin{document}

\title{The electromagnetic chirp of a compact binary black hole:\\
  a phase template for the gravitational wave inspiral}
\author{Zolt\'an Haiman$^{1,2}$}
\affiliation{
${}^{1}$Department of Astronomy, Columbia University, 550 W. 120th St., New York, NY, 10027, USA\\
  $^{2}$Department of Physics, New York University, New York, NY 10003, USA
}

\begin{abstract}
  The gravitational waves (GWs) from a binary black hole (BBH) with
  masses $10^4\lsim M\lsim 10^7{\rm M_\odot}$ can be detected with the
  {\em Laser Interferometer Space Antenna (LISA)} once their orbital
  frequency exceeds $10^{-4}-10^{-5}$ Hz.  The binary separation at
  this stage is $a=O(100)R_{\rm g}$ (gravitational radius), and the
  orbital speed is $v/c=O(0.1)$.  We argue that at this stage, the
  binary will be producing bright electromagnetic (EM) radiation via
  gas bound to the individual BHs. Both BHs will have their own
  photospheres in X-ray and possibly also in optical bands.
  Relativistic Doppler modulations and lensing effects will inevitably
  imprint periodic variability in the EM light-curve, tracking the
  phase of the orbital motion, and serving as a template for the GW
  inspiral waveform.  Advanced localization of the source by {\it
    LISA} weeks to months prior to merger will enable a measurement of
  this EM chirp by wide-field X-ray or optical instruments.  A
  comparison of the phases of the GW and EM chirp signals will help
  break degeneracies between system parameters, and probe a
  fractional difference difference $\Delta v$ in the propagation speed
  of photons and gravitons as low as $\Delta v/c \approx 10^{-17}$.
\end{abstract}

\pacs{04.30.-w, 04.25.-g, 04.80.Cc}
\date{\today \hspace{0.2truecm}}

\maketitle

\section{Introduction}

Advanced LIGO has detected gravitational waves (GWs) from three
stellar  binary black hole (BBH) mergers, GW150914, GW151226 and
LVT151012 \cite{Abbott_PRL_2016,Abbott_PRL_2_2016,LIGO_BBH} and
measured the GW inspiral waveform over the last $O(10)$ orbits.  While
no electromagnetic (EM) counterparts have been identified for these
events, such counterparts could exist for stellar-mass BBHs embedded
in dense gas, such as disks in active galactic nuclei (AGN)
\cite{Bellovary+2016,Bartos_2017,Stone_2017}.

The space-based GW detector LISA \cite{LISA} will be sensitive to
supermassive BBHs in the range $10^{4}-10^{7} \msun$ and is expected
to find dozens of merger events out to high redshifts ($z\sim 10$).
These BBHs are produced in mergers of galaxies, which deliver the
supermassive BHs, along with gas in the merging nuclei, to the center
of the merger-remnant galaxy.  The resulting compact BBHs are likely
embedded in a gaseous environment, producing bright EM emission.

Identifying a GW source in EM bands would have considerable payoffs
for cosmology and astrophysics \citep[e.g.][]{Phinney2009,Bloom+2009}.  Examples
for cosmology include measurement of the Hubble constant
\citep{Schutz_1986}, and probing gravity theories by comparing the
luminosity-distance relation inferred from photons and gravitons
\citep{DefMen07}.  Measuring the differences in the arrival times of
photons and gravitons from the same cosmological source would be an
independent test of the massive character of gravity and of possible
violations of Lorentz invariance in the gravity sector
\citep{KHM2008}.

In this paper, we focus on the last possibility.  While
previous work has suggested the idea of comparing the arrival times of
photons and gravitons \citep[e.g.][]{ceg01,ckr02,KHM2008}, a crucial
limitation has remained: in order to measure a difference in their
propagation speeds, one needs to know when the photons vs. gravitons
were emitted by the source.  In general, this would require modeling
the astrophysics of the source.  Here we argue that for a compact
massive BBH in the LISA band, such modeling is not required: as long
as the EM photospheres envelope individual BHs (rather than the binary
as a whole), the relativistic Doppler modulation from the orbital
motion of the binary will inevitably imprint the analog of the GW
chirp on the EM light-curve~\cite{DHS2015}\footnote{Ref.\cite{HayasakiLoeb2016} considered a similar
  Doppler modulation that would arise following the tidal disruption
  of a star by a massive BBH at somewhat larger separations, but did
  not consider simultaneous direct measurement of GWs.}. A measurement of
the EM light-curve will determine the phase of the binary's orbit.
Comparing the EM and GW chirp signals will help break degeneracies
between system parameters, and will measure the relative propagation
speed $\Delta v$ of photons {\it vs} gravitons.  Observing the same source with
{\em LISA} and a wide-field X-ray (and possibly optical) instrument
could probe a fractional difference as low as $\Delta v/c \approx
10^{-17}$.

\section{Periodic electromagnetic emission}

{\it EM emission from binaries --} Recent hydrodynamical simulations
have clarified the dynamics of a gaseous accretion disk around a BBH.
Although the binary torques the inner disk and creates a low-density
cavity, roughly twice the size of the binary separation, gas flows
into this cavity through narrow streams \cite{al96}.  Simulations in
which the BHs are in the computational domain have furthermore shown
that the individual BHs are fueled efficiently via their own
``minidisks''
\citep{Cuadra+2009,Roedig+11,Nixon+2011,Roedig+2012,Gold+2014,Farris+2014,Dorazio+2016}. This
allows merging massive BHs to remain as bright as a typical quasar,
nearly all the way to their merger \citep{Farris+2015}.

{\it EM emission from the BH components --} The gas accreting onto
each BH forms a minidisk extending out to its tidal truncation
radius.  In units of their respective gravitational radii, these are
$R_1/R_{\rm g1}\approx 0.27 q^{-0.3}(1+q)(a/R_{\rm g})$ and
$R_2/R_{\rm g2}\approx 0.27 q^{-0.7}(1+q)(a/R_{\rm g})$ for the
primary and secondary BH, respectively
\citep{Paczynski1977,al94,Roedig+2014}, where $a$ is the binary's
semi-major axis. Simulations resolving the minidisks have found good
agreement with these sizes \citep{Farris+2014}.  The nature of the
emission produced by accretion via each minidisk should resemble those
from quasars.

In standard quasar accretion disk models, the EM emission is
radially stratified, with high-energy radiation produced closer in.
In a steady, Keplerian, optically thick disk around a BH of mass
$M$, the thermal emission $\sigma_{\rm B} T^4$ at radius $R$ balances
the local dissipation $3GM\dot{M}/8\pi R^3$, where $T$ is the
effective temperature at radius $R$, $\dot{M}$ the accretion rate, and
$G$ and $\sigma_{\rm B}$ are Newton's constant and the
Stefan-Boltzmann constant~\citep[e.g.][]{frankbook}.

Integrating the black-body emission over the face of a disk with a
temperature profile of $T\propto R^{-3/4}$, the surface brightness
profile $I_\lambda(R)\propto [\exp((R/R_\lambda)^{3/4})-1]^{-1}$ has
the scale radius at the observed wavelength $\lambda_{\rm obs}$.
\begin{equation}
\frac{R_\lambda}{R_g}=68
  \left(\frac{\lambda_{\rm obs}}{0.55\mu{\rm m}}\frac{3}{1+z}\right)^{4/3}
  \left(\frac{M}{10^6 \msun}\right)^{-1/3}
  \left(\frac{\dot{M}}{\dot{M}_{\rm Edd}}\right)^{1/3}.
  \label{eq:R_lambda}
\end{equation}
Here $\dot{M}_{\rm Edd}\equiv L_{\rm Edd}/c^2$ is the accretion rate
corresponding to the Eddington luminosity $L_{\rm Edd}$, and we
adopted a radiative efficiency of $\epsilon=15\%$ and a quasar
luminosity of $L=\epsilon \dot{M} c^2 = 0.15 L_{\rm Edd}$.  The
fiducial wavelength of 0.55$\mu$m corresponds to the optical $V$ band;
in far UV bands ($\sim 0.15\mu$m) the disk would be $\sim 5$ times
more compact ($R_\lambda \sim 12 R_{\rm g}$).  Note that $\approx
20\%$ of the total flux at $\lambda_{\rm obs}$ arises from inside the
scale radius.

The X-ray emission observed from quasars is thought to arise from a
combination of thermal disk emission and a hot corona. The size of the
emitting region is limited by the broad observed FeK$\alpha$ line
width and from variability measurements to be $\lsim 10 R_{\rm g}$
\cite{ReynoldsNowak2003,MiniuttiFabian2004}. The X-ray emission from a
handful of quasars has been directly constrained via microlensing,
with half-light radii consistent with this compact size
\citep[e.g.][]{Dai+2010,Jimenez-Vicente+2015,Chartas+2016,Guerras+2017}.
These X-ray-emitting regions fit around the individual BHs until the
last stages of the merger, when the binary separation drops below
$\lsim 20 R_{\rm g}$.

Microlensing observations in the optical band have found sizes
consistent with eq.~(\ref{eq:R_lambda}). However, the observed
luminosities are significantly below the value predicted by the above
simple model, and imply $\sim3$ smaller disks
(e.g.~\cite{Morgan+2007}).  One resolution of this discrepancy is that
the optical/UV radiation emerging from the inner disk is scattered at
larger radii. For the compact binaries considered in this paper, the
individual minidisks are truncated at $(10-30)R_{\rm g}$. Reprocessing
may then occur only by the narrow streams of gas inside the
circumbinary cavity, with a small covering factor, and by the
circumbinary disk farther out.  We therefore consider the possibility
below that the optical/UV emission is generated at radii $\sim 3$
times smaller than eq.~(\ref{eq:R_lambda}).

{\it Periodic relativistic modulation --} The minidisks are fueled at
a rate that varies periodically, on timescales comparable to the
binary period
\citep[e.g.][]{Hayasaki+2007,MM08,Cuadra+2009,Roedig+11,Noble+2012,ShiKrolik:2012:ApJ,Dorazio+2013,Farris+2014,ShiKrolik2015,Dorazio+2016},
which may introduce a corresponding periodic variation in the thermal
EM emission~\cite{Farris+2015b}.  However, for compact binaries at
relativistic separations ($\beta\equiv v/c\gsim 0.1$, where $v$ is the
orbital velocity and $c$ the speed of light), strong periodic
variability is {\em inevitably} caused by relativistic Doppler
modulation, and is expected to dominate the variability for mass
ratios $q\equiv M_2/M_1\lsim 0.05$ \citep{Dorazio+2016}.  The Doppler
effect results from the orbital motion, and exists irrespective of the
details of the emission.

Photons emitted by gas bound to individual BHs suffer a Doppler shift
in frequency $D=[\Gamma(1-\beta_{||})]^{-1}$, where
$\Gamma=(1-\beta^2)^{-1/2}$ is the Lorentz factor,
$\beta_{||}=\beta\cos\phi\sin i$ is the component of the velocity
along the line of sight, with $i$ and $\phi$ the orbital inclination
and phase.  The apparent flux $F_\nu$ at a fixed observed frequency
$\nu$ is modified from the flux of a stationary source $F_\nu^{\rm 0}$
to $F_\nu =D^{3}F^{\rm 0}_{D^{-1}\nu}=D^{3-\alpha}F^{\rm 0}_{\nu}$.
The last step assumes an intrinsic power-law spectrum $F^{\rm
  0}_{\nu}\propto \nu^{\alpha}$.  To first order in $v/c$, this causes
a sinusoidal modulation of the apparent flux along a (circular) orbit,
by a fractional amplitude $\Delta F_\nu/F_\nu = \pm
(3-\alpha)(v/c)\sin i$~\footnote{For a circular orbit; non-sinusoidal
  modulations from eccentric orbits are easily incorporated into the
  analysis.}.

This Doppler modulation is a plausible explanation for the remarkably
sinusoidal optical and UV light-curves of the bright $z=0.3$ quasar
PG~1302-102~\cite{DHS2015}, originally discovered by
\citep{Graham+2015}.  This quasar, with a 5.2 yr observed period
($a\sim 0.01~{\rm pc}\sim300~{\rm R_g}$), is a mildly relativistic
binary BH candidate with $v/c\sim 0.07$.  For more compact binaries,
additional general relativistic effects from time-delays and lensing
will be important.  While we do not consider these effects here, we
note that they will enhance the periodic brightness modulations, while
tracking the phase of the orbital motion~\cite{Schnittman+2017}.

As long as the BHs have their own photospheres, a periodic brightness
modulation will be inevitable. The only exception would be the rare
case of either (i) a face-on binary, or (ii) a widely separated
equal-mass binary, inclined to the line of sight, for which the
lensing effects are negligible, and the blue-shift/brightening of one
BH cancels the red-shift/dimming of the other.


\section{Electromagnetic chirp for binaries in the LISA  band}


{\it LISA binaries --}
The key feature of the Doppler-induced periodic variability is that
its phase tracks the orbit of the binary.  For a massive BBH whose
inspiral is detected by LISA, this phase can be compared directly with
the evolving phase of the GW chirp signal.  The GW frequency (twice
the orbital frequency) is given by
\begin{equation}
  \label{eq:f_GW}
  f_{\rm GW} =
  1.4\times10^{-4}~{\rm Hz}~  \left(\frac{M}{10^6 \msun}\right)^{-1} \left(\frac{a}{60R_{\rm g}}\right)^{-3/2}.
 \end{equation}
The inspiral stage of a binary can be observed by LISA once its
observed frequency enters the LISA band ($f_{\rm obs}=(1+z)^{-1}f_{\rm
  GW}\approx 10^{-4}-10^{-5}~{\rm Hz}$ for $ M=10^4-10^7{\rm
  M_\odot}$).  A compact binary will inspiral and merge, due to GW
emission, on a timescale (to leading post-Newtonian quadrupole order;
\cite{Peters_1964}) of
\begin{equation}
  \label{eq:t_GW}
  t_{\rm GW} =
  0.17~\q^{-1}~{\rm yr}~   \left(\frac{M}{10^6 \msun}\right) \left(\frac{a}{60R_{\rm g}}\right)^4,
 \end{equation}
where $\q\equiv 4q/(1+q)^2$ is the symmetric mass ratio.  Binaries can
spend up to several years in the LISA band, and execute $\approx 1,900
\q^{-1}(a/60R_{\rm g})^{5/2}$ cycles, over which the GW chirp is
measured.

\begin{figure}
  \vspace{-5\baselineskip}
\centering\includegraphics[width=90mm]
{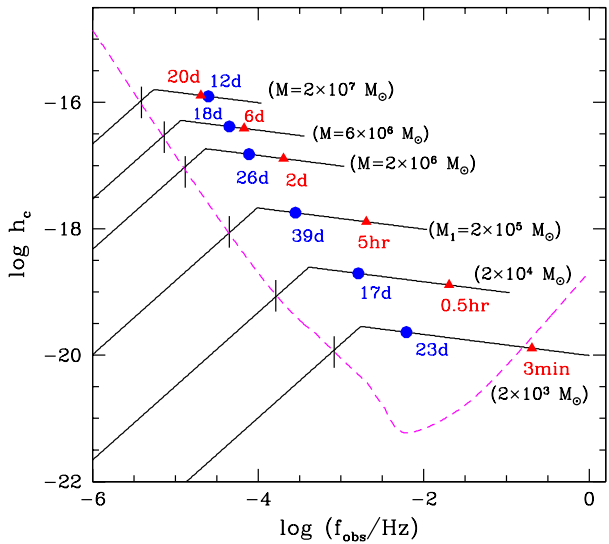}
\vspace{-7.\baselineskip}
\caption{\small\it The figure shows the tracks across the LISA band of
  equal-mass binaries with different masses, as labeled. The break in
  the characteristic strain $h_c$ marks 5 years prior to merger. Along
  each track, the marks correspond, from left to right, to the times
  when (i) the binary enters the LISA band (vertical line), (ii) the
  sky localization of a typical binary reaches an accuracy of 10
  deg$^2$ (blue circle), and (iii) the tidal truncation radius of the
  circumprimary disk becomes smaller than $10 R_{\rm g1}$ (red
  triangle).}
\label{fig:lisatracks}
\end{figure} 

The evolution across the LISA band of equal-mass binaries with several
different masses (all at $z=1$) is illustrated in
Figure~\ref{fig:lisatracks}.  The characteristic strain $h_c$, defined
as the product of the sky- and polarization-averaged Fourier strain
amplitude $h(f)$ and the square root of the number of cycles ($n_{\rm
  cycle}\equiv\min[f^2/\dot{f},f\tau]$, where $\tau=5$ years is the
assumed LISA mission lifetime) is shown, along with the spectral noise
density $S_{\rm n}(f)$. We adopted the LISA configuration with six
links, 2 million km arm length, and $\tau=5$yr (N2A2M5L6) from
ref.~\cite{Klein_2016}.

Along each track we mark, from left to right, three characteristic
times.  (i) First, equal-mass binaries with $(10^7, 10^6, 10^5)~{\rm
  M_\odot}$ enter the LISA band (defined as the time when the S/N
reaches $h_c/S_{\rm n}=1$) when their separation is (55, 110, 230)
$R_{\rm g}$, respectively (vertical marks).  (ii) As the binary
evolves and accumulates S/N, its parameters can be estimated to
increasingly better accuracy.  With the configuration we adopted, LISA
can localize a typical $10^6 \msun$ binary on the sky to an accuracy
of 10 deg$^2$ between a $\sim$week and a few months prior to the
merger, with localization degrading rapidly with redshift, to
$\sim$days at $z=3$ \citep{KHM2008,LangHughes2008,McWilliams+2011}.
The blue circle along each curve denotes the time when the
localization error drops to 10 deg$^2$ for a typical binary. For
simplicity, we assume that this occurs when the accumulated S/N=50;
this gives good agreement with the advance localization accuracies
found previously in \citep{KHM2008} and \cite{LangHughes2008}, who
adopted a similar noise density (the accuracies vary by an order of
magnitude, depending on the orientation of the orbital plane, spin
magnitude and direction, and sky position).  The advance localization
is important for beginning an EM monitoring campaign, and occurs at
separations of (16, 35, 69) $R_{\rm g}$.  (iii) Finally, the red
triangles denote the time when the tidal truncation radius of the
circumprimary disk becomes smaller than $10 R_{\rm g1}$.  This ensures
that the X-ray emitting regions of the disk have not yet been tidally
stripped, and occurs (20 days, 2.1 day, 5 hrs) prior to merger.  Note
that the typical $10^7~{\rm M_\odot}$ binary is localized too late,
but the $(10^6,10^5)~{\rm M_\odot}$ binaries are localized well before
this tidal stripping; they accumulate a total S/N=(214, 436) and
complete (232, 1510) cycles between localization and tidal stripping.
We list the evolution of binaries with a range of masses at $z=1$ and 2
in Table~1.

{\it Measuring the EM chirp --}
The advance localization will allow pointing a wide-field telescope at
the LISA error box and obtain a densely sampled EM light-curve,
covering hundreds of cycles.  We consider both X-ray and optical
telescopes, but note that because of the tidal truncation discussed
above, it is unclear if optical emission will still be present for a
binary in the LISA band.

In the optical, for a $10^6{\rm M_\odot}$ BH at $0.5<z<3$, shining at
the Eddington limit, the detection of 10\% variability will require a
sensitivity corresponding to 22-27 magnitudes~\cite{KHM2008}.  As an
example, the Large Synoptic Survey Telescope (LSST; \footnote{see
  \texttt{www.lsst.org}}) has a field of view (FOV) of 10 deg$^2$,
which can cover the typical 2D sky localization error box from LISA, a
month prior to merger, in a single pointing.  LSST can reach a
sensitivity of 27 mag (corresponding to a binary at $z=3$) in an
integration time of less than 1hr (see Table~1 in~\cite{KHM2008}).
This is shorter than the orbital period and allows constructing
a light-curve.

In the X-rays, as an example, the Wide Field Imager instrument on the
proposed Athena mission~\footnote{see
  \texttt{www.cosmos.esa.int/web/athena}} has a field of view of 0.5
deg$^2$ and an effective area of $\sim 1$m$^2$ at 1keV.  Tiling the
full LISA error box a month before merger would require $20$
pointings; alternatively, fewer cycles could be measured once the sky
localization improves (an accuracy of 0.5 deg$^2$ will be available
several days prior to
merger~\citep{KHM2008,LangHughes2008,McWilliams+2011}). The Lynx
mission currently being developed \footnote{See
  \texttt{wwwastro.msfc.nasa.gov/lynx}} has a similar capability.  For
a $10^6{\rm M_\odot}$ BH at $z=1$, with an X-ray luminosity of
$L_X=0.05 L_{\rm Edd}$, Athena would collect $\approx$10 photons in a
10$^3$ sec integration, i.e. would obtain a flux measurement to $\sim
30$\% accuracy throughout the 10 deg$^2$ area in $\sim$5
hrs. Continued tiling for a week would yield a measurement of $\sim$30
X-ray cycles throughout the field.

Once the periodic EM counterpart has been identified, the rest of a
month-long monitoring campaign can focus on this source, and can be
done by many other, smaller-FOV telescopes.  We conclude that
obtaining a well sampled EM light-curve should be feasible for
binaries at $z=1-2$ with masses ${\rm few}\times 10^3 {\rm
  M_\odot}\lsim M \lsim {\rm few}\times 10^6 {\rm M_\odot}$.  Less
massive binaries will have insufficient S/N, while more massive
binaries enter the LISA band when their separation is already too
compact for stable X-ray emission around the individual BHs (see
Fig.~\ref{fig:lisatracks} and Tab.~1).

\begin{figure}
\vspace{-5\baselineskip}
\centering\includegraphics[width=90mm]
{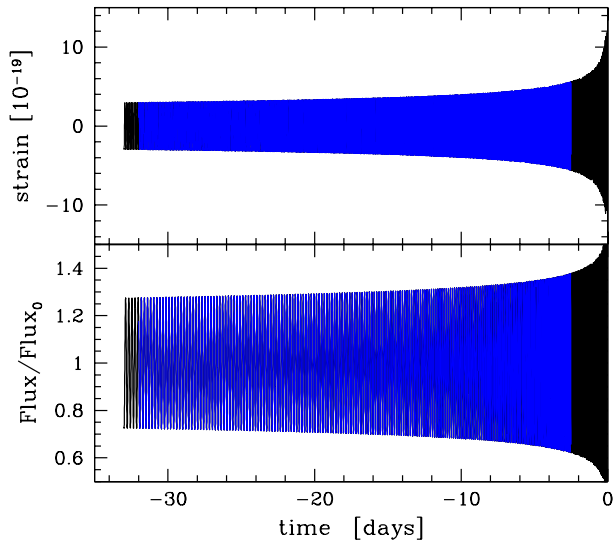}
\vspace{-7.\baselineskip}
\caption{\small\it Illustration of the time-domain GW chirp and the EM
  light-curve for an $M_1=10^6~{\rm M_\odot}$, $M_2=3\times 10^5~{\rm
    M_\odot}$, circular, edge-on binary at $z=1$.  For clarity, the EM
  light-curve shows only the Doppler modulation and exludes lensing
  effects. The part of the signal shown in blue marks the $\approx
  400$ cycles available for EM chirp measurement in X-rays, and
  possibly in other bands.  For this fiducial binary, the mean X-ray
  flux is $F_x=0.05L_{\rm Edd}/4\pi d_L^2=2\times10^{-15}~{\rm
    erg~s^{-1}~cm^{-2}}$.}
\vspace{-1.\baselineskip}
\label{fig:chirps}
\end{figure} 

\begin{figure}
\vspace{-5\baselineskip}
\centering\includegraphics[width=90mm]
{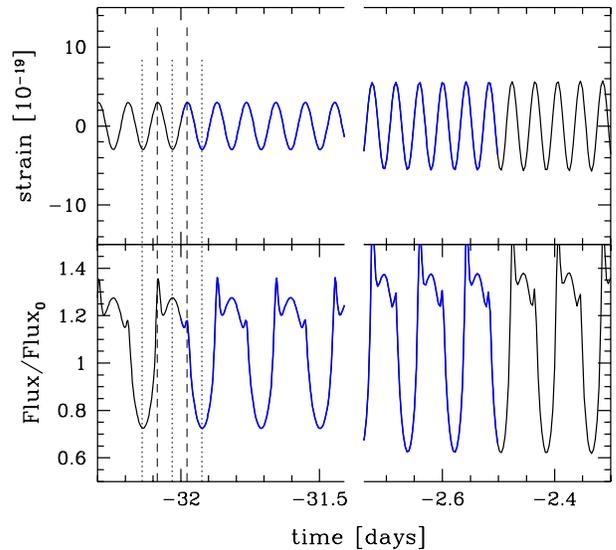}
\vspace{-7.\baselineskip}
\caption{\small\it This figure zooms in on the GW and EM chirp signals
  of the binary in Fig.~\ref{fig:chirps}, to show the few cycles
  around a look-back time of 32 days (when the binary is first
  localized on the sky to 10 deg$^2$) and around 2.5 days (when the
  circumprimary disk is tidally stripped down to 10 $R_{\rm g1}$). The
  bottom panel assumes a tilt of the orbital plane 10$^\circ$ away
  from edge-on, to illustrate the asymmetric lensing of the primary
  and the secondary (the sharp peaks marked by the dashed lines).}
\vspace{-1.\baselineskip}
\label{fig:chirpszoom}
\end{figure}


\section{Comparing the electromagnetic and gravitational wave chirp signals}


{\it Illustration --} Figure~\ref{fig:chirps} illustrates the
comparison between the EM and GW chirp signals.  The figure shows the
strain from an edge-on ($\cos i=0$), circular, $M_1=10^6{\rm
  M_\odot}$, $M_2=3\times10^5{\rm M_\odot}$ binary at $z=1$, with the
chirp $df/dt$ computed from the lowest-order PN (quadrupole) term.
The corresponding EM light-curve is shown assuming a spectral slope
$\alpha=-1$, modulated by relativistic Doppler boost at order $v/c$.
We have furthermore assumed that the secondary BH out-accretes the
primary, and is twice as luminous, based on the simulation results in
~\cite{Farris+2014}.  Note that if the spectral slope $\alpha$ remains
constant, the EM Doppler modulation amplitude ($v/c\propto f^{1/3}$)
increases less steeply with frequency than the GW strain amplitude
($\propto f^{2/3}$).  In principle, a few spectra, taken during the
inspiral, could determine any evolution of $\alpha$).  The figure
high-lights in blue the 387 cycles available for EM chirp measurement
in X-rays, and possibly in other bands (see Table~1).

Figure~\ref{fig:chirpszoom} zooms in on the GW and EM chirp signals of
the binary in Fig.~\ref{fig:chirps}, to show the few cycles around a
look-back time of 32 days (when the binary is first localized on the
sky to 10 deg$^2$) and around 2.5 days (when the circumprimary disk is
tidally stripped down to 10 $R_{\rm g1}$).  In this figure, we have
included the effect of gravitational lensing of whichever BH is behind
the other, using the standard formula for lensing magnification to
order $v/c$, $\mu=(x^2+2)/[x\sqrt{x^2+4}]$, where
$x\equiv\theta/\theta_E$ is the angular distance between the two BHs
on the sky, $\theta=\sqrt{(a\sin\theta)^2+(a\cos\theta\cos i)^2}d_A^{-1}$,
measured in units of the Einstein radius of the BH in the front,
$\theta_E\approx 2(R_{\rm gi}a)^{1/2}d_A^{-1}$.  Note that the effect is
asymmetric; to show this asymmetry, we have assumed a small 10$^\circ$
tilt of the orbital plane away from edge-on ($i=80$ deg).

Figure~\ref{fig:chirpszoom} shows several features of the
correspondance between the time-domain GW chirp and the EM
light-curve.  First, in this example of a nearly edge-on binary, the
GW is linearly polarized ($h_+$ only, $h_\times\approx 0$).  The
maxima of the GW strain correspond to the phase when the binary's axis
is aligned with the line of sight.  The Doppler modulation $\Delta
F/F=(3+\alpha)v_{||}/c$ at this phase vanishes, however, this also
coincides with the maximum lensing magnification.  The Einstein radius
for the primary is larger, causing a more pronounced brightening when
the secondary is being lensed (marked by the left vertical dashed
line), but the lensing of the primary is still visible (vertical
dashed line on the right).  The minima of the GW strain alternate to
sample the maxima and minima of the underlying Doppler modulation of
the EM light-curve (shown the three vertical dotted lines).

Clearly, the shape of the EM light-curve will depend on inclination
and mass ratio (see also~\footnote{D.J. D'Orazio and R. Di Stefano, in
  preparation}). Larger inclinations would eliminate the lensing
effect, but leave the phase of the EM chirp the same. On the other
hand, the GW strain would acquire a cross-component ($h_\times\neq 0$)
which is $\pi/2$ out of phase with the plus ($h_+$) component. Thus,
comparing the EM and GW phases amounts to an independent measurement
of the orbital inclination.  Note that in our examples, the binary
separation is $\gsim 20 R_{\rm g}$, but for smaller separations,
additional general relativistic effects, due to time-delay, and
higher-order lensing, will become important.

{\it Phasing accuracy --}
The signal-to-noise ratio $\rho$ for measuring a deviation in the GW
waveform is given by
\begin{equation}
  \rho^2(\delta h) = 2\times 4\int^{f_{\rm max}}_{f_{\rm
        min}}df\,\frac{|\delta h(f)|^2}{S_n^2(f)}
\label{eq:deltaSN}
\end{equation}
where $f_{\rm min}$ is the frequency when the observation begins
(i.e. when the localization accuracy drops below 10 deg$^2$) and
$f_{\rm max}$ is the frequency when the observation ends (i.e. when
the tidal truncation radius of the primary BH drops below $10 R_{\rm
  g1}$).  The extra factor of two on the right-hand side is because of
the assumed 3-arm configuration, equivalent to two independent
interferometers.  Because the EM chirp signal can provide a template
for the phase evolution, we assume that these phases are precisely
known, and evaluate the S/N for detecting a pure phase shift in the GW
waveform, $\delta h=h(f)(1-\exp[i \Delta\psi(f)])$. Here $h(f)=
\mathcal{A} f^{-7/6} \exp[i\psi(f)]$ is the frequency-domain inspiral
waveform in the stationary-phase approximation,
$\mathcal{A}=\pi^{-2/3}30^{-1/2}\mathcal{M}^{5/6} d_L^{-1}$ is the
amplitude in the leading Newtonian order, $\mathcal{M}\equiv
(M_1M_2)^{3/5}/(M_1+M_2)^{1/5}$ is the chirp mass, and $d_L$ is the
luminosity distance.

As a simple case, we consider a shift of the time-domain waveform by a
constant $\Delta t$, $\delta h=h(f)(1-\exp[2\pi i f \Delta t])$.  This
depicts a constant propagation speed difference
$\Delta v/c = c\Delta t/d_L$ of photons and gravitons, and
yields
\begin{eqnarray}
\label{eq:deltaSN-const1}
  \rho^2(\delta h) &=& 16\int^{f_{\rm max}}_{f_{\rm
      min}}df\,\frac{|h(f)|^2[1-\cos(2\pi f \Delta t)]}{S_n^2(f)}
  \\
&\approx& 8 (2\pi\Delta t)^2 \int^{f_{\rm max}}_{f_{\rm
      min}}df\,\frac{|h(f)|^2  f^2 }{S_n^2(f)},
  \label{eq:deltaSN-const2}
\end{eqnarray}
where in the last step we assumed $2\pi f \Delta t \ll 1$.
Fig.~\ref{fig:lisatracks} shows that most of the signal-to-noise for
$M\gsim 10^4~{\rm M_\odot}$ is contributed near $f_{\rm max}$, at the
highest frequency at which there is still X-ray emission.  To a good
approximation, a detectable shift corresponds to $\approx 10/\rho$
rad, for a source with overall signal-to-ratio $\rho$
\citep{will06,Kocsis_2011}.  Evaluating eq.~(\ref{eq:deltaSN-const1})
numerically, Table~1 lists the shift $\Delta v/c$ for each binary that
would produce $\delta\rho=10$.  We find values as low as $\Delta
v/c\approx 10^{-17}$.

If the velocity difference is due to gravitons with a nonzero mass
$m_g$ in a Lorentz-invariant theory~\cite[e.g.][]{HassanRosen2012},
then it will depend on frequency according to $\gamma m_g c^2 = h_p
f$, where $h_p$ is Planck's constant, $f$ is the GW frequency, and
$\gamma$ the Lorentz factor for the graviton.  If $m_g c^2 \ll h_p f$,
the graviton speed differs from $c$ by $\Delta v/c = 1/2 (m_g
c^2/h_p)^2 f^{-2}\propto f^{-2}$; this translates to a phase drift in
the GW waveform $\Delta\psi(f)\propto f^{-1}$~\cite{Will1998}.  In
Table~1, we lists the graviton mass $m_g$for each binary that would
produce $\delta\rho=10$, computed with this phase drift, using
eqs.~3.8 \& 3.9 in \cite{Will1998}.  We also list the equivalent
Compton wavelength ($\lambda_g=h_p/m_gc$).

Although Lorentz invariance for photons has been tested to high
accuracy, it could be violated in the gravity sector, especially over
cosmological scales.  This occurs, e.g. in brane-world scenarios with
extra dimensions \cite[e.g.][]{ckr02,ceg01}, or in vector-tensor
scenarios \cite[e.g.][]{jm04}; Lorentz-violating massive gravitons
have also been proposed as candidates for cold dark matter \cite[e.g,
][]{dub05}.  See refs.~\cite{bbw05,will06} for in-depth discussions.
Following the EM and GW chirp in tandem as they evolve over a decade
in frequency would explicitly test whether any measured phase drift is
consistent with the expectation from $\gamma m_g c^2 = h f$.

In principle, one should be able to tighten the above limits, by doing
a direct cross-correlation analysis, or matched filtering, between the
GW data stream and the light-curve of the EM counterpart(s).  We also
note that in case there are several variable quasars in the initial
LISA error box, then this analysis will also help identify the correct
counterpart to begin with (i.e. the object with period and phase
closely matching those of the GWs).

\begin{table*}[t]
\label{tab:numbers}
\centering
\begin{tabular}{c c |c  c  c c | c c  c  c  c  | c c c}
 & & \multicolumn{4}{c|}{at localization} & \multicolumn{5}{c|}{at tidal truncation} & \multicolumn{3}{c}{shift} \\
  \hline
  \hline
  $M_1$ & $M_2$& $a$ & $t_m$ & $f_{\rm min}$ & $\Delta F/F_0$ & $t_m$ & $f_{\rm max}$ & $\Delta F/F_0$ &  S/N & $N_{\rm cyc}$ & $\Delta v/c$ & $\log m_g$ & $\log\lambda_g$ \\
   $[{\rm M_\odot}]$ & $[{\rm M_\odot}]$ & [$R_{\rm g}$] & [day] & [Hz] & \% & [day] & [Hz] & \% & - & - & [log] & [eV/$c^2$] & [km]\\
  \hline
  \multicolumn{14}{c}{z=1}\\ 
 \hline
 $10^3$         & $10^3$    & 190 & 23 & -2.2 & 15 & 3min  & -0.7 & 46 & 64 & 19670  & $-17.5$ & $-24.0$ & 15.1 \\
 $10^4$         & $10^4$    & 99  & 17 & -2.8 & 20 & 0.5hr & -1.7 & 46 & 530 & 3895  & $-18.2$ & $-25.1$ & 16.2 \\
 \hline
 $10^5$         & $10^5$    & 69  & 39 & -3.6 & 25 & 0.2 & -2.7 & & 436 & 1510  & $-17.3$ & $-25.8$  &  16.9 \\
 $10^6$         & $10^6$    & 35  & 26 & -4.1 & 34 & 2.1 & -3.7 & 46 & 214 & 232   & $-16.1$ & $-26.3$  &   17.4 \\
 $3\times10^6$  &  $3\times10^6$  & 24  & 18 & -4.4 & 41 & 6.1 & -4.2 & 46 &87  & 58    & $-15.2$ & $-26.3$  &   17.4 \\
 $10^7$         & $10^7$    & 16  & 12 & -4.6 & 50 & 20  & -4.7 & & --  & --    & --      & --       & -- \\
 $10^5$         & $3\times10^4$  & 68  & 34 & -3.4 & 37 & 0.3 & -2.5 & 69 & 387 & 2044  & $-17.4$ & $-25.5$  &   16.6 \\
 $10^6$         & $3\times10^5$  & 38  & 32 & -4.0 & 50 & 2.5 & -3.5 & 69 & 195 & 387   & $-16.1$ & $-26.1$  &  17.2 \\
 $10^7$         & $3\times10^6$  & 18  & 15 & -4.5 & 74 & 25  & -4.5 & 69 & --  & --    & --      & --       & -- \\
 \hline
 \multicolumn{14}{c}{z=2}\\
 \hline
 $10^5$         & $10^5$    & 41  & 7.8 & -3.4 & 31  & 0.3 & -2.9 & 46 & 177 & 385   & $-17.1$ & $-25.5$ & 16.6 \\
 $10^6$         & $10^6$    & 22  & 6.6 & -4.0 & 42  & 3.1 & -3.9 & 46 & 60  & 37    & $-15.7$ & $-25.8$ & 16.9 \\
 $3\times10^6$  &  $3\times10^6$   & 16  & 4.7 & -4.2 & 51  & 9.2 & -4.3 & 46 & --  & --    & --      & --     & -- \\
 $10^5$         &  $3\times10^4$   & 40  & 5.9 & -3.2 & 49 & 0.4 & -2.7 & 69 & 145 & 456   & $-17.2$ & $-25.2$ & 16.3 \\
 $10^6$         &  $3\times10^5$   & 24  & 7.7 & -3.8 & 63 & 3.7 & -3.5 & 69 & 56  & 57    & $-15.8$ & $-25.7$ & 16.8 \\
  \hline
\end{tabular}
\caption{\small\it For binaries with different redshift, primary and
  secondary mass ($M_1$ and $M_2$; columns 1-2) the table shows the
  binary separation ($a$), time-to-merger ($t_m$), GW frequency
  ($f_{\rm min}$), and EM variability amplitude ($\Delta
  F/F_0=(3+\alpha)v_{||}/c$ with $\alpha=-1$) when the binary is first
  localized to 10 deg$^2$ (assumed to coincide with S/N=50;
  cols.~3-6); the time-to-merger ($t_m$), GW frequency $(f_{\rm
    max}$), EM variability amplitude ($\Delta F/F_0$), S/N, and \# of
  cycles accumulated since localization when the circumprimary disk is
  tidally truncated at $10 R_{\rm g1}$ (cols.~7-11); and a constant
  velocity difference ($\Delta v/c$) or graviton mass ($m_g$) and
  equivalent Compton wavelength ($\lambda_g$) which could be measured
  to S/N$=10$ (cols.~12-14).  }
\end{table*}

{\it Plasma effect for photons --}
The EM and GW signals can be offset from one another due to plasma
effects slowing down the photons.  The plasma frequency of the
intergalactic medium is proportional to the density of free electrons,
$n_e\approx 10^{-7}(1+z)^3{\rm cm}^{-3}$, yielding $\nu_p=(n_e e^2/\pi
m_e)^{1/2}=3~{\rm Hz} (1+z)^{3/2}$, and the corresponding index of
refraction $n=\left(1-\frac{\nu_p^2}{\nu^2} \right)^{1/2}\approx
1-10^{-34}\left(\frac{E}{1{\rm keV}}\right)^{-2}$.  This shows that
plasma effects for photons can become comparable to the $\Delta v/c$
measurement limit for radio waves with $\lambda\gsim 10$cm. This would
become relevant in searching for periodic Doppler modulation of radio
emission (e.g. from jets around the individual BHs, as proposed in
\cite{KulkarniLoeb2016}).

\vspace{-0.5\baselineskip}
\section{Discussion and conclusions}
\vspace{-0.5\baselineskip}

Constraining the graviton mass by comparing the GW waveforms and the
EM light-curves has been proposed previously in the context of compact
white dwarf binaries \cite[e.g.][]{chl03}, and the possibility of an
analogous measurement with massive BBHs was suggested by
\cite{KHM2008}.  Here we followed up on the latter suggestion, and
showed that the EM chirp signal, necessary to perform this experiment,
will inevitably be produced by the relativistic Doppler modulations of
the quasar-like emission from gas bound to the individual BHs.  We
also demonstrated that the EM chirp could be promptly (in a
$\sim$week) identified by a wide-field X-ray telescope, a $\sim$month
prior to merger, and would subsequently be available for monitoring
for 2-3 weeks by other, smaller-field-of-view telescopes.

The EM chirp will serve as a template for the GW inspiral, allowing a
novel test for theories with massive gravity or extra spatial
dimensions.  The constraints we find (Tab.1) are $\sim4$ orders of
magnitude tighter than the current upper limit,
$1.2\times10^{-22}~{\rm eV}/c^2$, from the LIGO waveforms
\cite{Abbott_PRL_2016}.  They are comparable to the sensitivity that
will be available from the phasing of GWs in LISA observations
alone~\cite{will06}.  Compared to the latter forecasts, we have used
only a restricted frequency-range, but have assumed that the EM
template breaks degeneracies between the phase drift and other system
parameters.  Additionally, the EM chirp template should be
useful in breaking parameter-degeneracies in the GW waveforms. As an
example, the sky position is correlated with the orbital
plane orientation; this degeneracy is broken by spin
precession~\cite{LangHughes2008} and by higher
harmonics~\cite{McWilliams+2011}, but only near the end of the
inspiral, when these effects are large.

As the inspiral proceeds from a binary separation of $\sim 60 R_{\rm
  g}$ to $20 R_{\rm g}$, the minidisks surrounding each BH will
gradually lose their outer annuli due to tidal stripping. Following
the evolution of the luminosity and spectrum of the source, during
this ``peeling off'' stage in the last few weeks of the inspiral, will
provide a novel tomographic probe of the radial structure of the
accretion disks.

We have conservatively focused here on the regime when the BHs are
separated by $\gsim 20 R_{\rm g}$, but the binary could possibly
remain bright past this stage, and exhibit periodic fluctuations even
during its last $\sim 10$ orbits~\cite[e.g.][]{Bode+2010}. However, it
is unclear whether the Doppler modulations, which arises from shocked
gas near the binary but bound to individual BHs, can be as
unambiguously related to the binary's orbital phase in this case as at
the larger separations we discussed.

We have focused on the X-ray emission, but binaries may be
bright at radio and gamma-ray energies (e.g. via Bremsstrahlung
emission and/or via jets \cite{Bode+2010,KulkarniLoeb2016}) all the
way up to the merger. It may then be possible to discover the
gamma-ray chirp signal without advance localization with an all-sky
monitor.  Barring this possibility, advance localization will
be necessary for this experiment to work, and LISA would need to
broadcast its data in real time.

While the gas accreting onto the BHs from a circumbinary
disk can create bright quasar-like emission, it is unlikely to
have an impact on the orbital evolution of the binary. When
the binary is in the LISA band, and inspirals on time-scale
of years, the gas torques are expected to be 4-5 orders of magnitude
weaker \cite{HKM2009,Kocsis_2011,Tang+2017}.

Future work will need to develop the method proposed here, by taking
into account effects we have neglected, such as the general
relativistic modulations during the binary's orbit, using higher-order
GW waveforms, as well as addressing the degeneracies between the full
set of system parameters.  The flux from quasars is known to vary
stochastically, including quasi-periodic modulations in the X-rays
bands. In the case of binaries, significant X-ray emission can arise
from shocks in accretion streams and the circumbinary disk, which may
be modulated periodically, on time-scales different from those of the
binary's orbit~\cite{Roedig+2014,Farris+2015b}.  The uncertainty to
which the periodic Doppler modulations can be measured, in the
presence of this variability, as well as realistic measurement errors
of the EM light-curves, will need to be assessed in future work, as
well.

\begin{acknowledgments}
  I thank Alessandra Buonanno, Csaba Cs\'aki, Daniel Chung, Kohei
  Inayoshi, Feryal Ozel, Lorenzo Sironi, and Luigi Stella for useful
  discussions, and Daniel D'Orazio, Bence Kocsis and Geoffrey Ryan for
  useful comments on a draft of this manuscript. I also gratefully
  acknowledge support by a Simons Fellowship in Theoretical Physics
  (ZH) and by NASA grant NNX15AB19G.
\end{acknowledgments}


%

\end{document}